\documentclass[a4paper,11pt]{article}
\pdfoutput=1 

\usepackage{jinstpub} 

\usepackage{lineno}

\title{\boldmath Belle II status and prospects for studies of neutral currents}

\author{V. Bertacchi}
\affiliation{Aix Marseille Univ, CNRS/IN2P3, CPPM, Marseille, France}

\emailAdd{bertacchi@cppm.in2p3.fr}

\abstract{
The Belle~II experiment at the SuperKEKB energy-asymmetric electron-positron collider is a substantial upgrade of the B factory facility at the Japanese KEK laboratory. Belle~II collected a sample of $362~\mathrm{fb}^{-1}$ at the $\Upsilon(4S)$ resonance between 2019 and  2022, with a maximum peak luminosity of $4.7~\times 10^{34}~\mathrm{cm}^{-2}s^{-1}$. Belle~II is currently facing a long shutdown period, required for several upgrades of the detector and the collider. Data taking will resume at the end of 2023. We report the recent measurements which involve neutral current transitions in $B$ meson decays. In particular, we present the current status and future prospects for the branching fractions measurements of the radiative decays $B\to K^*\gamma$ and the fully inclusive $B\to X_s\gamma$,   the search for $B^+\to K^+\nu\overline \nu$ decays,  the measurement of the branching fractions of $B\to J/\psi(\to \ell\ell)K$ and $B\to K^*\ell\ell$.  Finally, we show the perspectives of the search for $B\to K^{*}\tau\tau$ and the searches of lepton flavor violating channels $B\to K^{(*)}\ell\ell'$,  with $\ell=e, \mu, \tau$. 
}


\arxivnumber{arXiv:2307.04928}

 \collaboration{\includegraphics[height=17mm]{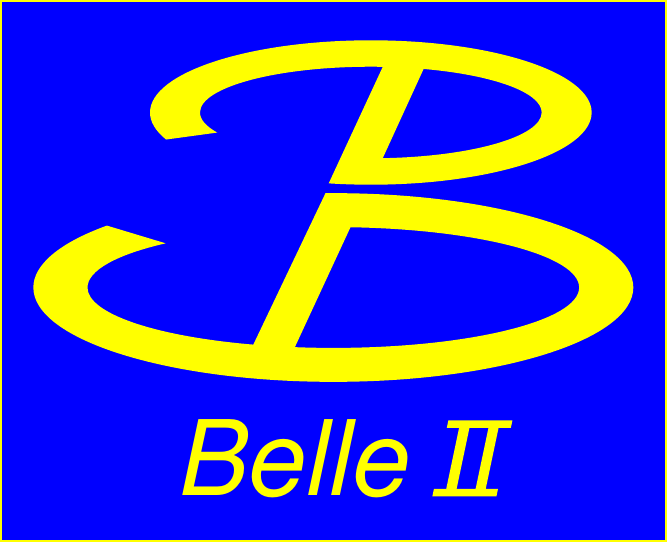}\\[6pt]
On behalf of Belle~II Collaboration}

\proceeding{New Frontiers in Lepton Flavor\\
  15-17 May 2023\\
  Pisa}

\notoc
\begin{document}
\maketitle
\flushbottom
\section{Introduction}

The  flavour changing neutral current $b\to s$ transitions are suppressed in the Standard Model (SM) and therefore sensitive to Beyond the Standard Model (BSM) amplitudes. The SM branching fractions are ${\mathcal O(10^{-5}-10^{-7})}$, predicted with ${10-30\%}$ uncertainties. Angular distributions and ratios can be used to improve the precision and eventually have access to new physics properties.

 Belle~II~\cite{Belle2:TDR} and SuperKEKB~\cite{SuperKEKB:TDR} produce an optimal environment to study the neutral currents.  
Belle~II has similar and good performance in electron and muon channels, in term of efficiency, fake rate and particle identification capability. This is a key feature to perform lepton flavour universality (LFU) tests and lepton flavour violation (LFV) searches in the $b\to s\ell\ell^{(')}$ sector, where $\ell$ indicates a charged lepton. On the other hand, the $b\to s\gamma$ and $b\to s\nu\overline \nu$ transitions represent a unique opportunity for Belle~II, because of the almost complete hermeticity of the detector, the possibility to exploit the $\Upsilon(4S)$ initial state constraint and the relatively low combinatorial background of the SuperKEKB collisions. 

One of the key tools of Belle~II for the channels with missing energy in the final state is the B-tagging, a set of reconstruction techniques to identify the  ${\Upsilon(4S)\to B\overline B}$ events exploiting the initial state knowledge to constrain the missing information in the signal side. It consists in reconstructing the partner $B$ meson, called $B_\mathrm{tag}$, produced in association with the signal one, to infer the properties of the signal. 
We refer to hadronic or semileptonic tagging according to the channels used for the $B_\mathrm{tag}$ reconstruction. 
  The B-tagging algorithm is called Full Event Interpretation (FEI)~\cite{Keck:FEI}, a boosted decision tree (BDT)-based tagging algorithm which exploits  a hierarchical approach to explore and reconstruct $\mathcal{O}(10^4)$ decay chains on the $B_\mathrm{tag}$ side. The efficiency for the hadronic (semileptonic) tag is 0.5\% (2\%) with a purity of 30\% (10\%).

\section{Fully inclusive \texorpdfstring{$B\to X_s\gamma$}{B to Xs gamma}} 

We present the measurement the $B\to X_s\gamma$ branching ratio as a function of photon energy in the range $1.8~\mathrm{GeV}<E_\gamma<2.7~\mathrm{GeV}$, where $X_s\gamma$ is the inclusive final state involving a photon and a strange hadron.  The measurement is performed on a $189~\text{fb}^{-1}$ Belle~II sample~\cite{Belle2:XsGamma}. 
The decays are reconstructed using the  hadronic $B$-tagging,  requiring a $\gamma$ in the signal side with a threshold energy of 1.4~GeV. The main challenge of the analysis is to suppress the background without breaking the inclusivity of the measurement.  The backgrounds are suppressed using a BDT and the residual $X_d$ background is estimated using simulated events. 
The signal is extracted by fitting the tag side $M_{\rm bc}=\sqrt{E^{*2}_\mathrm{beam}-p^{*2}_B} $ distribution  (where $ E_\mathrm{beam}^*$ is the beam energy and $p_B*$ is the  $B$ meson momentum in the center-of-mass frame), as a function of $E_\gamma$. The result is competitive with previous measurements performed with hadronic  $B$-tagging~\cite{Babar:Xsgamma}. The result in term of the partial branching fraction as a function of the photon energy is shown inf Fig.~\ref{fig:plots} (Left). 

The prospects of this measurements with larger statistics depend on the chosen photon energy threshold~\cite{Belle2:snowmass}. With lower threshold the background will be higher, while with higher threshold the theoretical  uncertainties will be higher. However, some improvements are expected both on background suppression side and  by using additional tagging methods, which will allow to reach a percent level precision~\cite{Belle2:snowmass}.  Measurements of relative quantities, such as asymmetries, will allow for further reduction of systematic effects.

\section{Measurement of  \texorpdfstring{$B\to K^*\gamma$}{B to K* gamma} branching fractions}

We present the measurement of the branching fraction of $B\to K^*\gamma$, where $K^*$ indicates both $K^{*+}(892)$ and $K^{*0}(892)$. The measurement is performed on a  $63~\text{fb}^{-1}$ Belle~II sample~\cite{Belle2:Kstargamma}. 
The decays are identified reconstructing only the signal $B$ in the event. The misreconstructed $\gamma$ background is suppressed with an energy selection, and with a veto on $\gamma$ from $\pi^0$ and $\eta$ decays. The $e^+e^-\to q \overline q$ background is suppressed with a multivariate analysis. The misreconstructed $K^*$ background is suppressed using the $K^*$ helicity angle distribution. A fit to $\Delta E=E_B^*-E_\mathrm{beam}^*$ (where   $E_B$ is the energy of the $B$ meson) is used to extract the signal, excluding higher-mass $K^*$ resonances. The results are $\mathcal B(B^0\to K^{*0}(K^+\pi^-)\gamma)=(4.5\pm 0.3\pm 0.2)\times 10^{-5}$, $\mathcal B(B^0\to K^{*0}(K_S^0\pi^0)\gamma)=(4.4\pm 0.9\pm 0.6)\times 10^{-5}$, $\mathcal B(B^+\to K^{*+}(K^+\pi^0)\gamma)=(5.0\pm 0.5\pm 0.4)\times 10^{-5}$, $\mathcal B(B^+\to K^{*+}(K_S^0\pi^+)\gamma)=(5.4\pm 0.6\pm 0.4)\times 10^{-5}$, where the first uncertainty is statistical and the second systematic, compatible with the world averages~\cite{PDG}.

This measurement is performed as the cleanest exclusive channel in $B\to X_s\gamma$ sector, and it is a first step toward asymmetry measurements of radiative decays. In the latters several systematic uncertainties cancel out, and projections based on Belle result~\cite{Belle:Kstargamma} shows that a precision below the percent level can be reached with few $\mathrm{ab}^{-1}$~\cite{Belle2:snowmass}.

\section{Search for  \texorpdfstring{$B^+\to K^+\nu\overline \nu$}{B+ to K+ nu nubar} decays}

The search of ${B^+\to K^+\nu\overline \nu}$ decay is a unique opportunity for Belle~II. This decay has never been observed before the amplitude~\cite{theo:Knunu} can receive sizeable contribution from BSM amplitudes. The measurement is performed on a sample with an integrated luminosity of  $63~\mathrm{fb}^{-1}$~\cite{Belle2:Knunu}. The reconstruction is performed with an inclusive tagging approach,  reconstructing the $B_\text{sig}$ using the highest $p_T$ track compatible with a $K^+$, and assigning the rest of the event to the $B_\text{tag}$. The procedure is validated on ${B^+\to J/\psi(\to \mu\mu)K^+}$ decays. Two  BDT in cascade are used to suppress the background exploiting the event shape, kinematics and vertex features. 

\begin{figure}[htbp]
\centering 
\includegraphics[width=.45\textwidth]{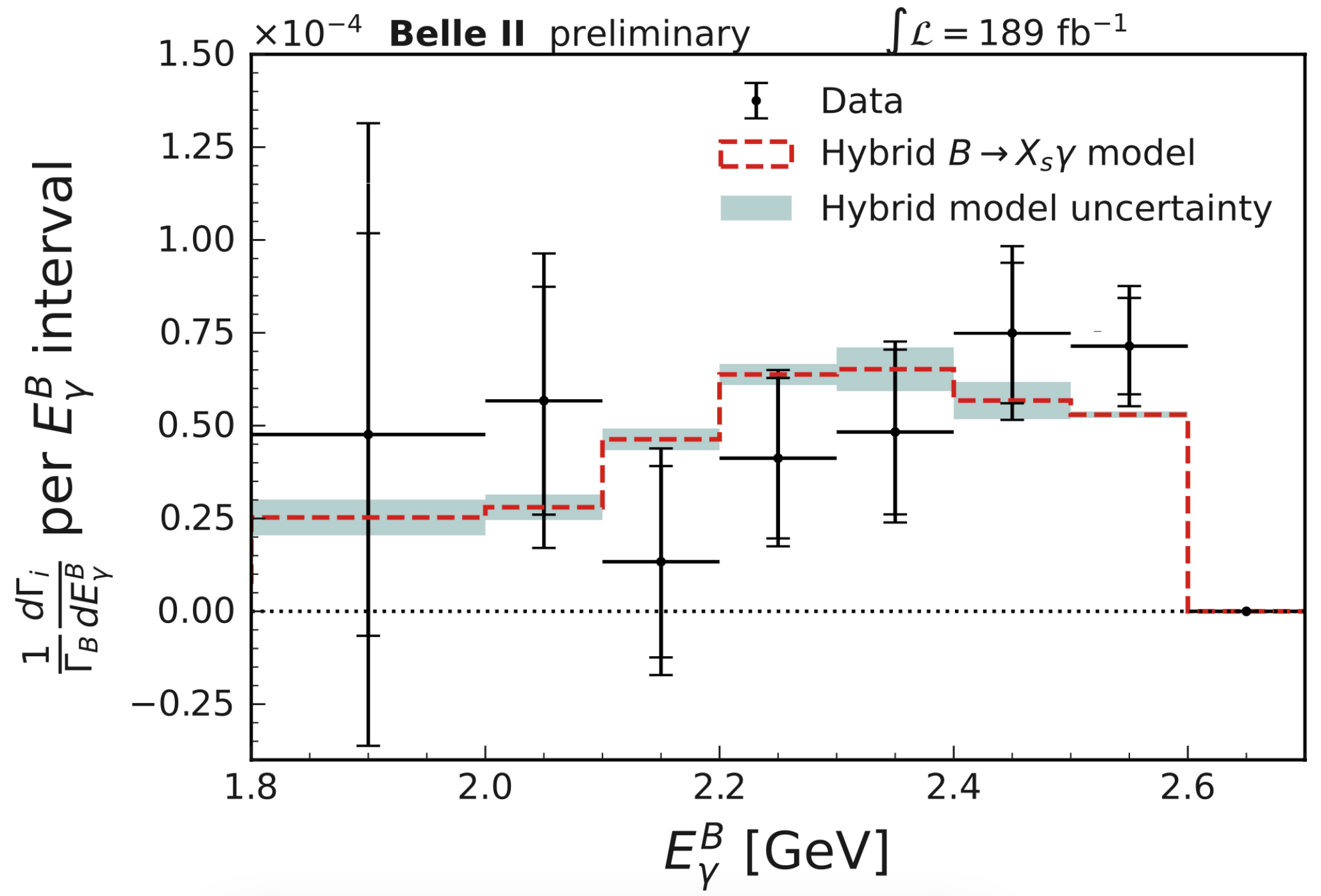}
\qquad
\includegraphics[width=.48\textwidth]{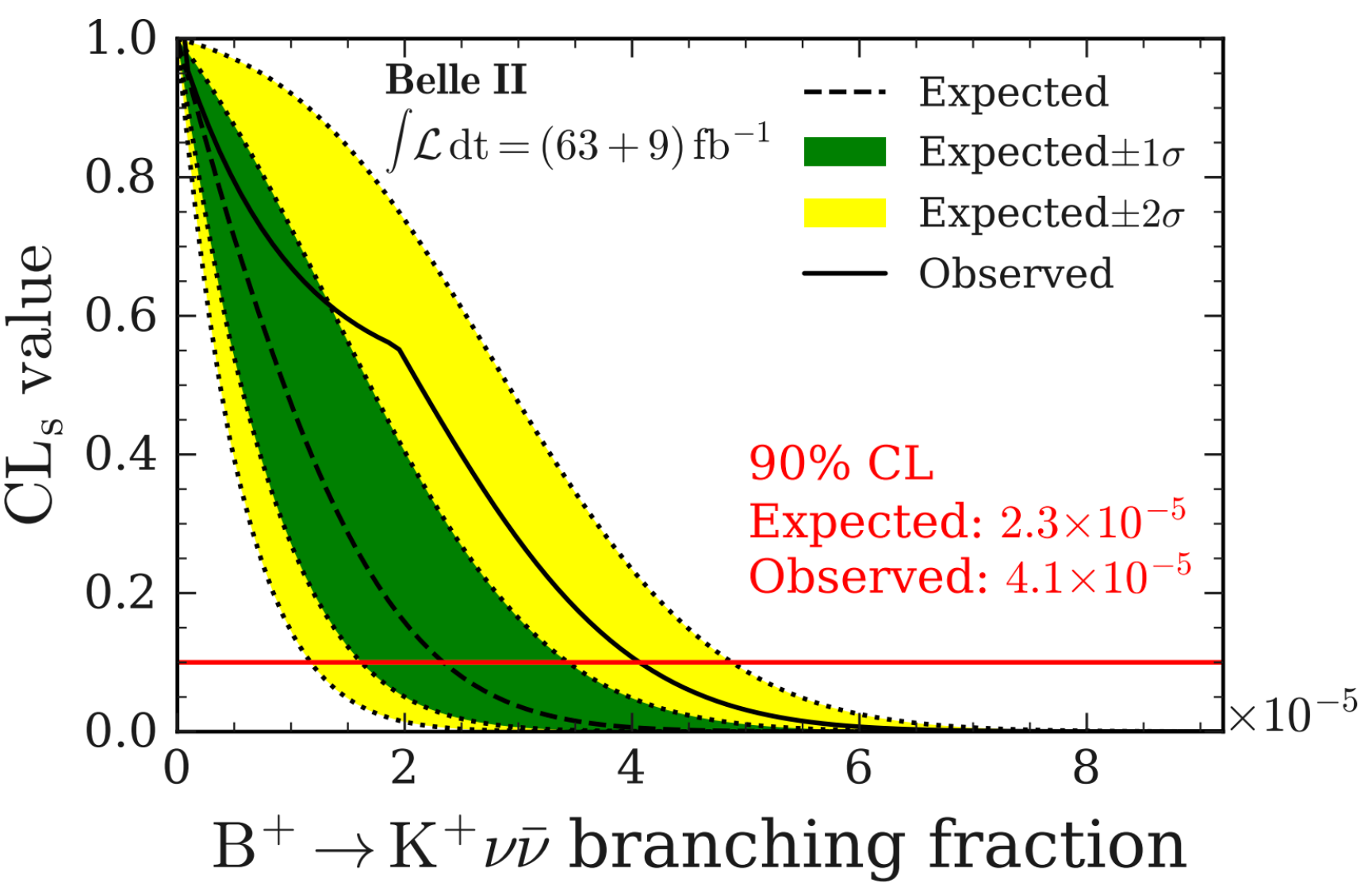}
\caption{\label{fig:plots} Left: partial branching fraction $(1/\Gamma_B)(d\Gamma_i/dE_\gamma)$ as a function of $E_\gamma$; the outer (inner) uncertainty bar shows the total (statistical) uncertainty~\cite{Belle2:XsGamma}. 
Right: $\mathrm{CL}_\mathrm s$ as a function of $\mathcal{B}(B^+\to K^+\nu\overline \nu)$ for the expected and observed signal yields, together with the corresponding 90\% confidence level upper limits. The expected limits are evaluated with background-only hypothesis~\cite{Belle2:Knunu}. }
\end{figure}

No signal is observed and the result is shown in Fig.~\ref{fig:plots} (Right) in term of upper limit.  This correspond to  ${\mathcal B( B^+\to K^+\nu\overline \nu)=(1.9\pm 1.3\,\text{(stat)}^{+0.08}_{-0.07} \,\text{(syst)})\times 10^{-5}}$, compatible with the SM prediction and the previous results~\cite{PDG}. 

The projection with larger samples~\cite{Belle2:snowmass} shows that a $5\sigma$ observation can be achieved with an integrated luminosity of $5~\mathrm{ab}^{-1}$ with an expected 50\% efficiency improvement coming from the use of exclusive tagging approaches in combination with the inclusive one. Moreover, additional channels ($K^*, K_S^0$) will be investigated.

\section{Measurement of  \texorpdfstring{$R_{K}(J/\psi)$}{RK(J/psi)}}

We present the measurements of the branching fraction of $B\to J\psi(\to \ell\ell)K$, $\ell=e,\mu$ and $K=K^+, K_S^0$, performed on a $189~\mathrm{fb}^{-1} $ Belle~II sample~\cite{Belle2:RJpsi}. The ratios $R_K(J/\psi)=\mathcal B(B\to J\psi(\to \mu^+\mu^-)K)/\mathcal B(B\to J\psi(\to e^+e^-)K)$ are also measured. These channels have no sensitivity on BSM,  so $R_K(J/\psi)\approx 1$ is expected. This analysis is used to validate the measurement of $B\to K^*\ell\ell$. The yields are extracted from a fit to  ${(\Delta E, M_{bc})}$ distribution. The results are $R_{K^+}(J/\psi)=1.009\pm0.022\pm0.008$, $R_{K_S^0}(J/\psi)=1.042\pm0.042\pm.008$, where the first uncertainty is statistical and the second systematic, in agreement with the expectations.

\section{Measurement of  \texorpdfstring{$B\to K^*\ell\ell$}{B to K* l l} branching fractions}

We present the measurement of the branching fractions   ${\mathcal B(B\to K^* \ell^+\ell^-)}$ (where ${\ell=e,\mu}$ and ${K^*=K^{*+}(892)}$, ${K^{*0}(892)}$) performed on a sample with an integrated luminosity of $189~\text{fb}^{-1}$~\cite{Belle2:Kll}. The backgrounds are suppressed using a BDT combined with a veto on the dilepton invariant mass for the ${J/\psi,\psi(2S)\to \ell\ell}$ background. An extended maximum likelihood fit is performed to the ${(\Delta E, M_{bc})}$ distribution.  The results are  $\mathcal B( B\to K^*\mu^+\mu^-)=(1.19\pm 0.31^{+0.08}_{-0.07})\times 10^{-6}, \mathcal B( B\to K^*e^+e^-)=(1.42\pm 0.48\pm 0.09)\times 10^{-6}, \mathcal B( B\to K^*\ell^+\ell^-)=(1.25\pm 0.30^{+0.08}_{-0.07})\times 10^{-6}$,
where the first uncertainty is statistical and the second systematic, compatible with the world averages~\cite{PDG}.

These results prepare the ground for the measurement of $R_{K^{(*)}}=\mathcal B(B\to \mu^+\mu^-K^{(*)})/\mathcal B(B\to e^+e^-K^{(*)})$, which will require a larger sample.

\section{ \texorpdfstring{$B\to K^*\tau\tau$}{B to K* tau tau} perspectives}

The measurement of $B\to K^*\tau\tau$ is complementary to the previously discussed searches, investigating the new physics in the third generation. The SM branching ratio is $\mathcal O(10^{-7})$, but BSM amplitudes can enhance the signal of several order of magnitude~\cite{Theo:Ktautau}. Currently the decay has been never observed and an upper limit at $\mathcal O(10^{-3})$ has been set~\cite{Belle:Ktautau}.  Prospects extrapolating from the current upper limit with a larger samples show that Belle~II can investigate the branching ratios down to $10^{-4}$ with $5~\mathrm{ab}^{-1}$, using hadronic and semileptonic B-tagging and reconstructing the $\tau$ leptons both in leptonic and hadronic decays~\cite{Belle2:snowmass}.

\section{Perspectives of  Lepton Flavor Violation searches in  \texorpdfstring{$B\to K^{(*)}\ell\ell'$}{B to K(*) l l'} sector}

Several measurement have been performed in past years by BaBar, LHCb and Belle collaborations in the $B\to K^{(*)}\ell\ell'$ sector, where $\ell=e,\mu,\tau$, setting upper limits that span from $10^{-5}$ to $10^{-9}$ level. Belle~II is planning to join the effort in the searches of new physics in this sector.  Focusing on  $B\to K^{(*)}\tau\ell$,  the use of the hadronic or semileptonic tag allow to avoid the explicit reconstruction of the $\tau$ lepton. The signal is extracted from the $\tau$ recoil mass distribution, obtained from the $B_\mathrm{tag}$ and the signal $K$ track information. The recent results performed on Belle sample using the FEI are very promising~\cite{Belle:KtauEll}.




\acknowledgments

This project has received funding from the European Union’s Horizon 2020 research and inno- vation programme under the ERC grant agreement No 819127.

\bibliographystyle{JHEP}
\bibliography{biblio.bib}
\end{document}